\documentclass[%
 reprint,
superscriptaddress,
 amsmath,amssymb,
 aps,
prb,
]{revtex4-1}

\usepackage{graphicx}
\usepackage{dcolumn}
\usepackage{bm}
\usepackage{braket}

\begin{document}

\preprint{APS/123-QED}

\title{Minimal Ingredients for Orbital Texture Switches at Dirac Points in Strong Spin-Orbit Coupled Materials}

\author{J.A. Waugh}
\author{T. Nummy}
\author{S. Parham}
\affiliation{Department of Physics, University of Colorado at Boulder}
\author{Qihang Liu}
\author{Xiuwen Zhang}
\author{Alex Zunger}
\affiliation{Renewable and Sustainable Energy Institute, University of Colorado at Boulder}
\author{D.S. Dessau}
\affiliation{Department of Physics, University of Colorado at Boulder}
\affiliation{Renewable and Sustainable Energy Institute, University of Colorado at Boulder}

\date{\today}

\begin{abstract}
Recent angle resolved photoemission spectroscopy measurements on strong spin-orbit coupled materials have shown an in-plane orbital texture switch at their respective Dirac points, regardless of whether they are topological insulators\cite{qt2} or ``trivial'' Rashba materials\cite{labios22}. This feature has also been demonstrated in a few materials (\(\text{Bi}_2\text{Se}_3\), \(\text{Bi}_2\text{Te}_3\), and BiTeI) though DFT calculations \cite{labios22,qt2,qt3}. Here we present a minimal orbital-derived tight binding model to calculate the electron wave-function in a two-dimensional crystal lattice. We show that the orbital components of the wave-function demonstrate an orbital-texture switch in addition to the usual spin switch seen in spin polarized bands. This orbital texture switch is determined by the existence of three main properties: local or global inversion symmetry breaking, strong spin-orbit coupling, and non-local physics (the electrons are on a lattice). Using our model we demonstrate that the orbital texture switch is ubiquitous and to be expected in many real systems. The orbital hybridization of the bands is the key aspect for understanding the unique wave function properties of these materials, and this minimal model helps to establish the quantum perturbations that drive these hybridizations.
\end{abstract}

\maketitle

\section{Introduction}
In quantum systems the key piece of information that describes the physics involved is the Hamiltonian and the wave functions of the system’s constituents. Typically we are interested in the electron energies, momenta, and spin states, i.e. the electronic band structure. However, with the recent interest in strongly spin-orbit coupled systems and topological materials, it is becoming clear that there is additional critical information, i.e. that pertaining to the orbital wavefunctions and symmetries, their relative phases, and how they couple with the spin degrees of freedom of the material.  

In 3D Topological Insulators (TI’s) the inversion of an odd number of bands per unit cell leads to the necessity of a topological surface state with Dirac-like dispersion, and a frequently-described momentum-locked helical spin-structure that is left-handed above the Dirac point and right-handed below \cite{qt3, qi2011topological}. However, such a description ignores the fact that the J states and not the spin states are the relevant eigenstates of the spin-orbit coupled system, so there must be a richer manifold of entangled spin and orbital states than described in this simplistic picture. This was shown by detailed ARPES experiments in a prototypical TI \(\text{Bi}_2\text{Se}_3\) \cite{labios22, qt3, qt7, qt8}. As part of this physics, different orbital states directly couple to specific spins \cite{qt9}. In the case of \(\text{Bi}_2\text{Se}_3\), the different orbitals can couple to spins that do not follow the net helicity of the spin bands \cite{labios22, qt2, qt7}. This has ramifications when considering the bands as being entirely spin polarized, since in reality they are a superposition of opposing spins coupled to different orbitals.

A similar situation exists for Rashba states, in which the conventional picture is the spin-split parabolic band\cite{labios3, qt6}. In this picture the electron wavefunction is simply the two split bands with the spin component pointing in opposing directions. The opposite spins couple to the magnetic field (or broken inversion symmetry at the surface) and raise or lower the electrons energy. More precisely, a recent work \cite{qt2, qt12} has shown a complicated orbital and spin texture that is highly reminiscent of that of the TIs. In particular, there are spins of both helicities in the inner and outer Rashba bands, and these spins may couple to orbitals of different types. Elucidating the origin and underlying symmetry requirements for the spin behavior and especially the orbital texture switch is the goal of the present paper.

In an earlier work \cite{liu2016orbital} we used DFT to study the effects of spin-orbit induced hybridization in multi-band solids, including both topological insulators with band inversion as well as Rashba bulk solids. In that work we showed quite generally that SOC-induced hybridization of different azimuthal orbital momenta leads to a truncation of the spin magnitude in each band below its maximal value of \(\pm1\), with different levels of spin truncation in different bands arising from different orbital textures in those bands. Distilling the minimal ingredients that drives such physics is, however, difficult to access from these DFT calculations.

Here, we use an orbitally-intuitive minimal model of Rashba states at the outset, both for solving the electronic structure problem and for explaining the crucial couplings responsible from the main effects, focusing especially on the crucial orbital-texture switch, which has been observed to occur exactly at the Dirac points. 

We show that the orbital texture switch is determined by the existence of three main properties: local or global inversion symmetry breaking, strong spin-orbit coupling, and non-local physics (the electrons are on a lattice). Using our model we demonstrate that the orbital texture switch is ubiquitous and to be expected in many real systems. The orbital hybridization of the bands is the key aspect for understanding the unique wave function properties of these materials, and this minimal model helps to establish the quantum perturbations that drive these hybridizations.

\section{Model}
We shall model Rashba bands with a tight binding model of a two-dimensional sheet of hexagonal or square lattice atoms. Each site will have the atomic states \(p_x\), \(p_y\) and \(p_z\) orbitals centered on them, where the z-axis is perpendicular to the plane of atoms. We chose to neglect s-orbitals because they have no angular momentum, therefore not contributing to spin-orbit coupling. The basis set chosen is that used by many other DFT projections in the field \cite{qt7}. Since there is strong spin-orbit coupling, the basis set cannot be separated into spin and orbital components separately. The basis set must instead contain a full set of spins and orbitals assuming there is coupling of each orbital to any arbitrary spin. In order to account for this, the basis is
\begin{equation}
	\ket{ p_x, \sigma_z^+ },    \ket{ p_x, \sigma_z^-},   \ket{ p_y, \sigma_z^+ }, \ket{ p_y, \sigma_z^-} , \ket{ p_z, \sigma_z^+}, \ket{ p_z, \sigma_z^-}
\end{equation}
Where \(p_i\) are p orbitals in the 3 Cartesian directions (i=x,y,z), and \(\sigma_z\) is the spin component in the out of plane direction. 

We take the Hamiltonian from Peterson and Hedergard \cite{qt6}.
\begin{equation}
	H_0 = \sum{t_{\alpha\beta}(R_i-R_j) \ket{p_\alpha (R_i),\sigma} \bra{p_\beta (R_j) ,\sigma }}
\end{equation}
Where
\begin{align}
	&t_{\alpha\beta}(R_i-R_j)  \notag\\ &=\begin{cases}
		\omega \cos^2\theta - \delta sin^2 \theta & \text{for } (\alpha,\beta) = (x,x)\\
		(\omega - \delta) \cos \theta \sin \theta & \text{for } (\alpha,\beta) = (x,y) \text{ or } (y,x) \\
		\omega \sin^2 \theta - \delta \cos^2 \theta & \text{for } (\alpha, \beta) = (y,y) \\
		\gamma \cos \theta & \text{for } (\alpha,\beta) = (x,z) \text{ or } (z,x) \\
		\gamma \sin \theta & \text{for } (\alpha,\beta) = (z,y) \text{ or } (y,z) \\ 
		-\delta & \text{for } (\alpha,\beta) = (z,z)
	\end{cases}
\end{align}
We then add spin orbit coupling in the atomic basis form:
\begin{equation}
	H = H_0 + H_\text{SOC} 
\end{equation}
with
\begin{align}
	H_\text{SOC} =& \frac{\alpha}{2} L\cdot S  \notag \\=& \frac{\alpha}{2} 
	\left( \begin{array}{cccccc}
		0&0&-i&0&0&1\\
		0&0&0&i&-1&0\\
		i&0&0&0&0&-i\\
		0&-i&0&0&-i&0\\
		0&-1&0&i&0&0\\
		1&0&i&0&0&0	
	\end{array}\right)
\end{align}
Where the basis set is \(\ket{ p_x, \sigma_z^+ },    \ket{ p_x, \sigma_z^-},   \ket{ p_y, \sigma_z^+ }, \ket{ p_y, \sigma_z^-} , \ket{ p_z, \sigma_z^+}, \ket{ p_z, \sigma_z^-}\)

The \(\gamma\) term in the Hamiltonian allows for the hopping of an electron from an in plane p orbital to a neighboring atom’s \(p_z\) orbital, and in this simplest 2D model will only be present if there is an out-of-plane distortion or field. In a bulk 3D system this will usually come from a surface term (the classic Rashba effect) though it can also come from an intrinsic symmetry-breaking field or distortion \cite{labios3, labios4, qt10,labios2}. 
\begin{equation}
	\gamma = \bra{p_z(R)} V(z) \ket{p_n(R+x)}, (n=x,y)
\end{equation}
This hopping shows up in the Hamiltonian as off-diagonal elements between in-plane and \(p_z\) orbitals. These hopping elements of the Hamiltonian develop a momentum dependence, having no interaction at k=0 (Gamma point). These terms mix the basis states further than just the off-diagonal SOC terms, which are k-independent.

This entire Hamiltonian has some symmetries by design. First is the crystal symmetry, chosen here as either a hexagonal, rectangular or square lattice. This lattice allows for the electrons to hop, therefore bringing in a non-local momentum-dependence despite being built out of localized atomic orbitals. Next, there is out-of-plane inversion symmetry breaking, i.e. the \(\gamma\) terms. Lastly, there is spin-orbit coupling in the form explained previously. As we will show, it is with the combination of all three of these ingredients that we produce the unique orbital texture switch observed in the experiments and the DFT calculations.  Other interesting features of these states such as ``backwards'' and/or ``partial'' spin polarization are also readily duplicated and understood using these simple terms.

\section{Model Solution}
\begin{figure*}
	\centering
	\includegraphics[width=0.9\linewidth]{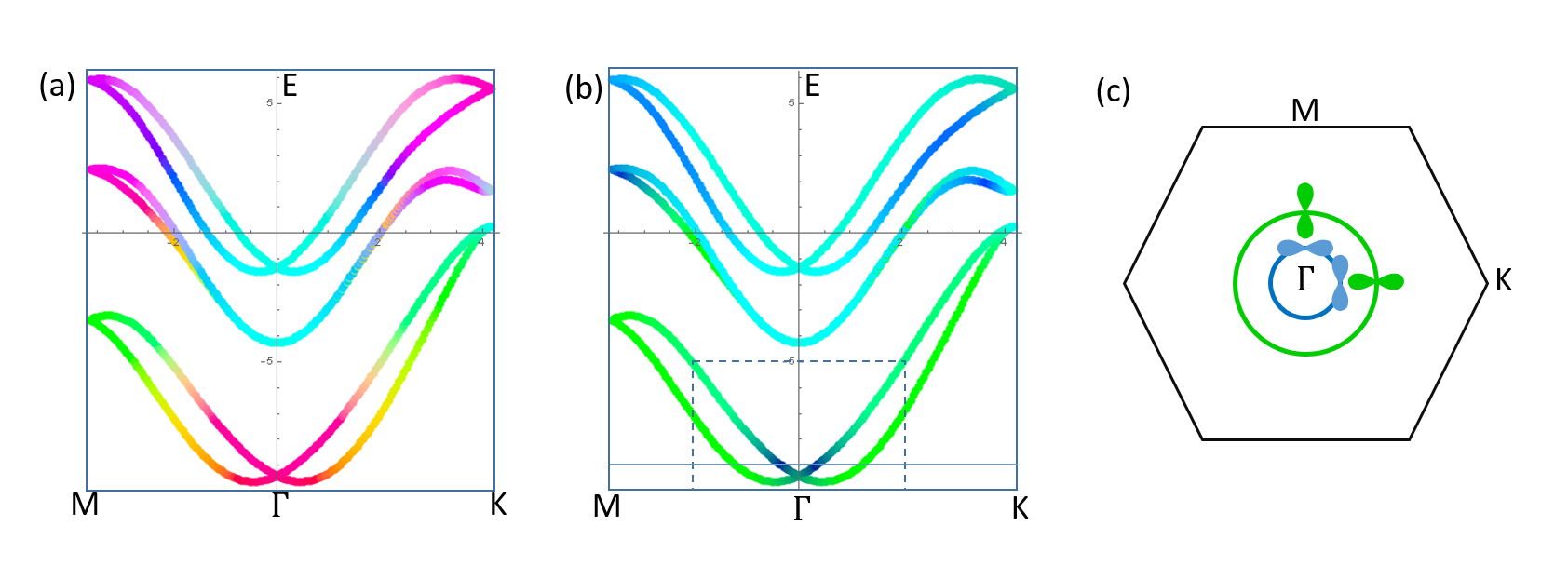}
	\caption[Band dispersion solution to the model Hamiltonian]{\label{fig1_tomo}(a) Band dispersion solution to the model Hamiltonian. The bands are colored according to their orbital contribution, giving a (R,G,B) color at each point corresponding to (\(p_z\), \(p_{rad}\), \(p_{tan}\)) contribution. Thus, a red point corresponds to \(p_z\) dominated, while green and blue points correspond to radial and tangential p-orbital dominated respectively. A teal color would correspond to equal parts radial and tangential. (b) shows the same band structure with the red (\(p_z\)) component turned off in the coloring. This shows the underlying orbital texture switch in the lowest Rashba band pair. (c) Shows the Bruillion zone and the two Rashba bands at the energy shown by the horizontal line in figure (b). Additionally, the dominant in-plane orbital contribution is shown on each rashba Band.   }
\end{figure*}

The Hamiltonian of equation 4 has three main components: orbital hybridization \(\omega\) and \(\delta\), spin orbit coupling \(\alpha\), and the out-of-plane symmetry breaking field \(\gamma\).  Figure \ref{fig1_tomo} shows the band solution to the model Hamiltonian with parameter choices \(\alpha=-2.5\), \(\delta=1.5\), \(\omega=0.5 \), and \(\gamma=1\)  that are reasonable for a typical ``strong'' spin-orbit compound on a hexagonal lattice. 6 bands are observed, as equation 1 begins with a 6 state basis. All bands, but especially the lowest pair of bands, exhibit a typical Rashba-like band structure corresponding to an ``inner'' and ``outer'' set of bands that are degenerate at \(\Gamma\). All bands are made of a combination of various orbitals and spins, with the mixing ratios of the spins and orbitals determined when the Hamiltonian is diagonalized. The coloring of the bands in both panels \ref{fig1_tomo}a and \ref{fig1_tomo}b indicate the orbital decomposition of the wavefunctions, including all three orbitals (panel \ref{fig1_tomo}a) or the in-plane orbitals only (panel \ref{fig1_tomo}b). The upper four bands have principally in-plane character (\(p_x\), \(p_y\) or \(p_{rad}\), \(p_{tan}\)) at the gamma point (blue/green), while the lower two bands have principally out-of-plane character at the gamma point (\(p_z\) or red). Ignoring minor splittings these would nominally correspond to the \(J_{3/2}\) states (two upper branches) and the \(J_{1/2}\) states (lower branch), though from the diagrams it is clear that this nomenclature is only reasonable near the zone center. 

\begin{figure*}
	\centering
	\includegraphics[width=0.4\linewidth]{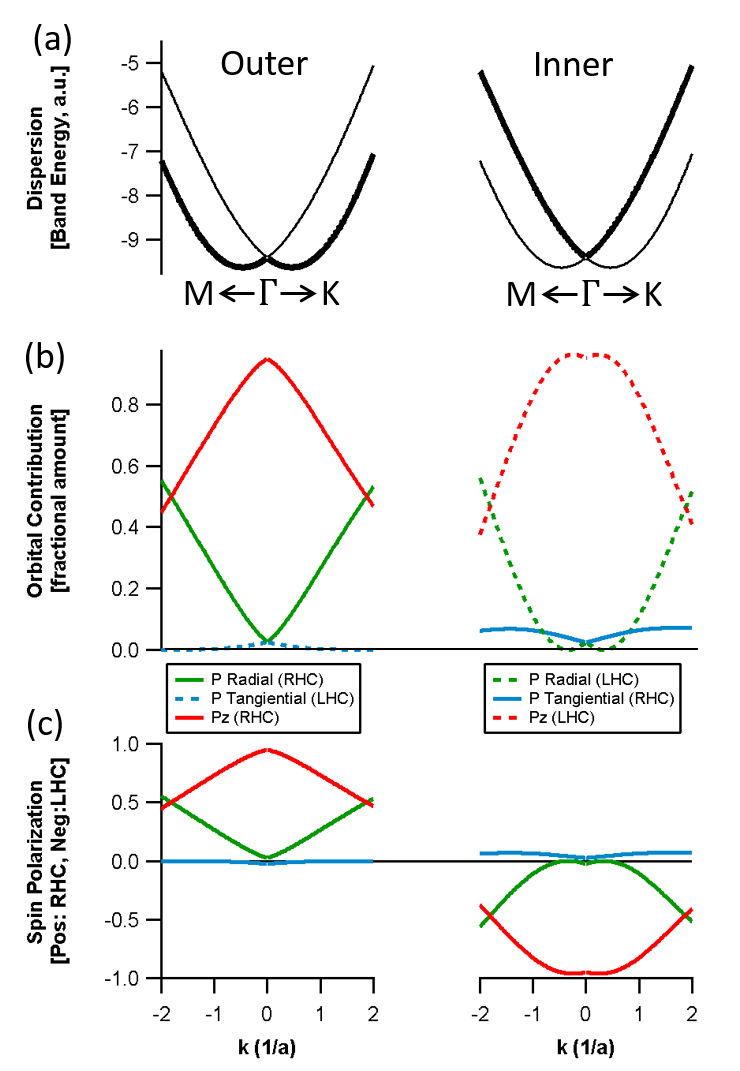}
	\caption[Orbital and spin decomposition]{\label{fig2_tomo} (a) The band dispersion of the same lowest set of Rashba bands from Figure 1 (see dashed box on figure 1(b)). Here, we separate the inner and outer Rashba bands. On the left column the outer bands are highlighted, and the right column the inner bands are highlighted. (b) The orbital contribution of the bands. The red lines show the \(p_z\) component, showing how they dominate at gamma point, and decrease in strength when moving away. The green shows the radial component, which has an overall trend of increasing while moving away from the Gamma point. It however shows a distinct difference on the inner Rashba bands where the weight decreases to zero before increasing. The tangential component decreases in the outer bands, but increases in the inner bands when moving away from Gamma. (c) The spin of these orbital contributions. Both the radial and out of plane p orbitals have a right handed spin on the outer band and a left handed spin on the inner bands. The tangential p orbitals have opposing spin helicity.  }
\end{figure*}

Figure \ref{fig2_tomo} shows more details of the orbital and spin contributions of the lower pair of Rashba-split states near the zone center, over the k-space range shown by the rectangular box in figure \ref{fig1_tomo}b. The left panels of figure \ref{fig2_tomo} show the breakdown for the outer states (bold, panel \ref{fig2_tomo}a) and the right panels show the breakdown for the inner states. It can be seen from panel \ref{fig2_tomo}b that at Gamma the \(p_z\) orbital (red) dominates the wave function of both inner and outer states, though this dominance quickly decays as one moves away from the gamma point. Additionally, we can see that at Gamma, the radial and tangential orbitals have a small and equal contribution to the wavefunction. As we move far away from Gamma the radial component quickly grows, and the tangential and out of plane components decrease. In the inner bands, the tangential component initially raises in contribution, while the radial component initially decreases. This is the fundamental aspect of the orbital texture switch in these Rashba bands – one band picks up a radial contribution while the other picks up a tangential one. Next, as it applies to spin (fig \ref{fig2_tomo}c), we can see that in the outer bands, both the out-of-plane and the radial components have right handed helicity, while the tangential component carries a left handed spin. In the inner bands the situation is reversed and the radial and out of plane components carry a left handed spin, and the now stronger tangential bands carry a right handed spin. An important aspect here is that the \(p_z\) and radial states carry the same spin helicity, while the tangential states carry an opposite helicity, with all helicities switching when going from the inner to the outer Rashba band.  This is identical to the situation discovered empirically for the Dirac state in the TI’s \(\text{Bi}_2\text{Se}_3\) and \(\text{Bi}_2\text{Te}_3\) \cite{labios6}, though here we show how it comes directly from a simplistic model.

The superposition of these opposing helicities in these bands can create unique spin polarizations, and reduce the overall net magnitude of spin measured in experiments. This has been an issue in many Topological Insulator experiments, and we demonstrate here that this feature should be expected to be present in nearly all Rashba materials (even if the effect is small). For the case of carefully selecting the light’s electric fields to be in the plane of the material, it is possible to ignore the out-of-plane orbital in the photo-emission process, and therefore measure the spin of purely these in-plane orbitals \cite{labios22}. These in-plane orbitals have spin components that oppose each other, giving rise to complete control over the photoelectron spin. By coming in with normal incidence light (E-field in the plane of the sample so selecting only in-plane orbital states), and changing the polarization from linear horizontal, vertical, +sp, -sp, +circular, and –circular, it should be possible to controllably and reproducibly eject photoelectrons with their spin along any arbitrarily chosen direction (x,y,z or anywhere in-between). This as a technically feasibility has been demonstrated multiple times in recent ARPES measurements \cite{qt8,qt9}.

\begin{figure*}
	\centering
	\includegraphics[width=0.8\linewidth]{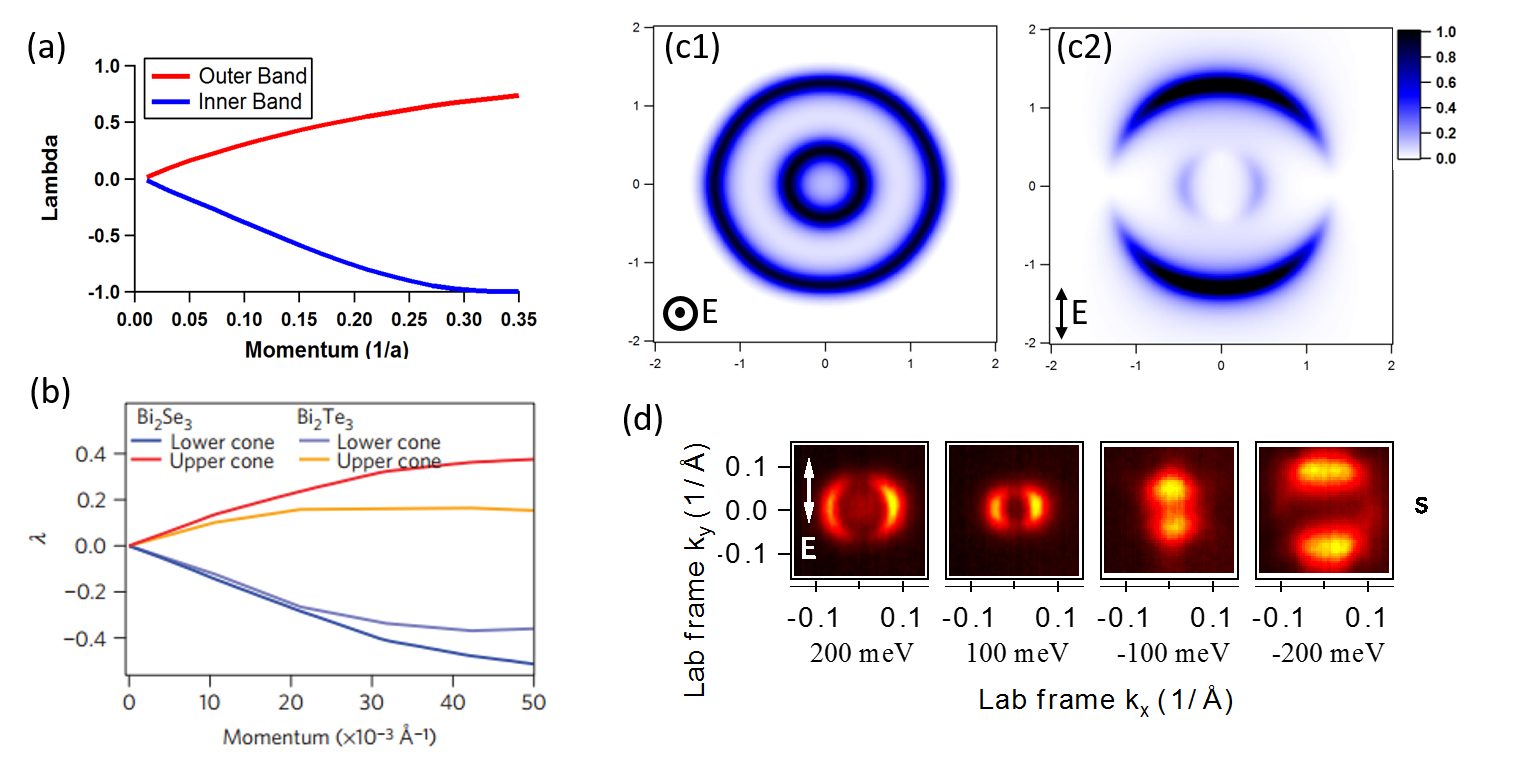}
	\caption[Comparason to ARPES on $\text{Bi}_2\text{Se}_3$]{\label{fig3_tomo}(a) Orbital asymmetry parameter lambda as a function of crystal momentum away from the gamma point for the Rashba bands calculated here. The outer band has a positive lambda indicating predominantly radial character to the in-plane states, while the inner band is negative, indicating predominant tangential in-plane character.  (b) the same lambda plot calculated from first principals for the surface Dirac states of the topological insulators \(\text{Bi}_2\text{Se}_3\) and \(\text{Bi}_2\text{Te}_3\) (taken from \cite{labios22}). The effect is very similar for the Rashba (a) and TI materials (b), though the magnitude of the effect (strength of lambda) is greater for the Rashba case. (c1) simulated ARPES spectrum for p-polarized light on the Rashba bands, showing both inner and outer band at the same constant energy slice. (c2) simulated ARPES spectrum for s-polarized light, highlighting the orbital texture switch by showing the nodes in spectral weight changing from being along the kx=0 axis to the ky=0 axis when going from outer to inner band. (d) Experimental constant energy cuts of \(\text{Bi}_2\text{Se}_3\) taken with s-polarized light, from \cite{labios22}. Shown here is the same structure as seen in (c2), switching from left-right dominated at higher energies (inner bands) to top-bottom dominated at lower energies (outer bands).   }
\end{figure*}

Figure \ref{fig3_tomo} compares another aspect of this Rashba simulation with calculations and experimental data from the prototypical 3D topological insulator, \(\text{Bi}_2\text{Se}_3\). We can characterize the strength of the orbital polarization through the orbital polarization parameter \(\lambda\), originally defined for the TI’s \(\text{Bi}_2\text{Se}_3\) and \(\text{Bi}_2\text{Te}_3\) in ref \cite{labios22}: 
\begin{equation}
	\lambda = \frac{I_0(|k|) - I_{90}(|k|)}{I_0(|k|) + I_{90}(|k|)}
\end{equation}
where \(I_0\) and \(I_{90}\) are the photo-emission intensities along two orthogonal high-symmetry directions when using properly polarized incident photons, or equivalently, the projected orbital polarizations. Figure \ref{fig3_tomo}(a) shows the k dependence of the \(\lambda\) term for the two lower bands, inner and outer, in the Rashba system calculated here, while figure \ref{fig3_tomo}(b) shows the k-dependence of \(\lambda\) for the upper and lower Dirac cones in \(\text{Bi}_2\text{Se}_3\) and \(\text{Bi}_2\text{Te}_3\) calculated from DFT projected intensities. Clearly the trends of the two systems are extremely similar, with the main difference that the model Rashba system has a more dramatic in-plane orbital texture switch (with magnitude approaching unity) than the TI’s, which have maximum magnitude of approximately 0.5. Additionally, we can simulate an expected ARPES spectrum of these Rashba bands. As expected, we see two concentric circles in k-space at a constant energy surface if we come in with p-polarization (selecting out of plane orbitals). However, if we instead come in with a light polarization in the plane of the material, we select the in-plane orbitals, and see arcs of ARPES intensity which have opposing directions. The outer Rashba band shows arcs top-bottom, while the inner one shows arcs left-right. This can be compared directly to the measurement of \(\text{Bi}_2\text{Se}_3\) reproduced in figure \ref{fig3_tomo}(d), which shows that the upper Dirac cone exhibits the left-right arc pattern, while the lower Dirac cone exhibits a top-bottom spectral intensity pattern.

\begin{figure*}
	\centering
	\includegraphics[width=0.6\linewidth]{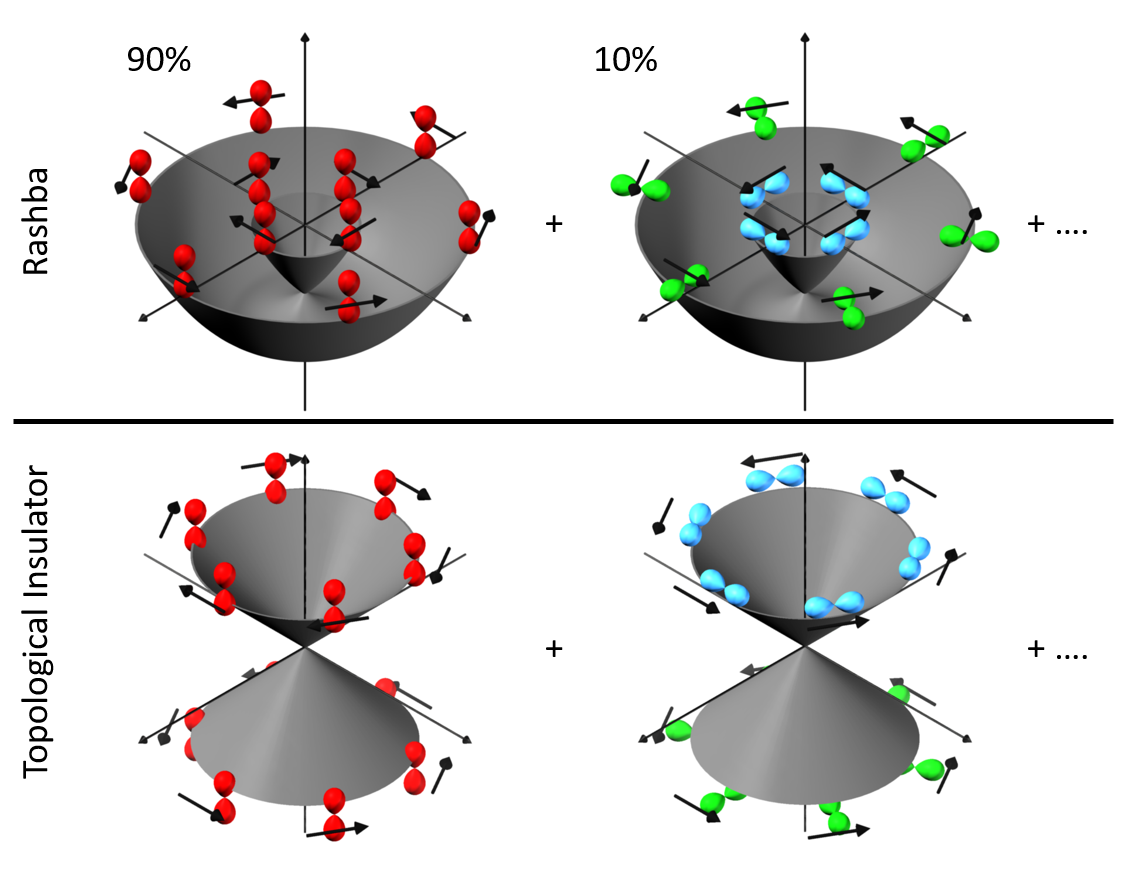}
	\caption[Cartoon showing spin-orbital decomposition]{\label{fig4_tomo} Cartoon showing the band structure of the lower set of Rashba bands (upper panel) and surface Dirac bands in Topological Insulators (lower panel, reproduced from reference \cite{labios22}). With a simple mapping of Inner Rashba to Upper Dirac cone we observe a dramatic similarity in all aspects of the orbital and spin makeup of these bands. In particular, the \(p_z\) orbitals (green – left panels) dominate these bands and have a left/right spin helicity upon crossing the degeneracy point. The weaker in-plane orbital components (right panels) have right-handed spin helicity on both sides of the degeneracy point and show an in-plane orbital texture switch from predominantly tangential to predominantly radial.  }
\end{figure*}

Figure \ref{fig4_tomo}c and \ref{fig4_tomo}d show cartoons that summarize our findings for both the nominal \(J_{1/2}\) Rashba bands (top) and surface Dirac bands from the TI compounds \(\text{Bi}_2\text{Se}_3\) and \(\text{Bi}_2\text{Te}_3\). It can be seen for both materials that the bands actually built out of a superposition of orbitals. Shown in the cartoon, they are composed of 90\% out-of-plane p orbitals with a 10\% contribution of in-plane orbitals (coupled to their own spins). The out-of-plane orbitals couple to the traditional spin helicity expected in both Rashba and TI bands. Separately, the in-plane orbitals actually couple to spin in a unique fashion, giving a right handed spin texture to both the inner and outer Rashba bands (upper and lower Dirac cones). The orbitals themselves are also not uniquely radial or tangential, and in fact, switch their dominance at the Gamma point in both materials. For the Rashba bands the inner band is dominated by tangential p orbitals close to the gamma point, while the outer band is dominated by radial orbitals.

\section{Discussion}
The inversion symmetry breaking term \(\gamma\) expanded to first order in crystal momentum k shows a linear dependence near gamma. This term hybridizes the in-plane and \(p_z\) orbitals, breaking the usual assumption where the bands would not interact at all. This interaction term additionally can cause an avoided crossing in the band structure, where at the anti-crossing the two bands are strongly hybridized and demonstrate the most mixing. 

When spin-orbit coupling is turned on, the degeneracy of the bands is lifted due to further mixing among each spin state. These spin states, however, are also coupled to orbital angular momentum, so the orbitals themselves must also mix. It is through this coupling that the bands can develop unique features such as orbital texture switches centered around various high symmetry points. The most striking of these is at the Gamma point, where the orbital texture switches from a radial to tangential texture, having direct consequence to experiments on the materials. 

\section{Conclusion}
It is also possible to extend this model to non-2-dimensional materials as well. By further extending the model in the standard tight binding approach, it will be possible to calculate the orbital texture of bands in materials with more complicated atomic bases. These materials may show unique orbital textures for each atom type in the material, as the basis would be a summation of p orbitals on each like-atom in the material. This will help further understand experiments that are sensitive to the depth of the material. 

Here we presented a simple model with few restrictions: local or global inversion symmetry breaking, strong spin-orbit coupling, and non-local physics (electrons on a lattice). This model shows the orbital texture switch seen in ARPES studies\cite{labios22,qt2,qt7}. This model also predicts that this feature is not unique to these materials, and in fact should be present in all strong-spin orbit coupled materials with broken inversion symmetry. This would suggest that this feature is as ubiquitous as the classic spin splitting seen in spin-orbit coupled models. This feature is also not restricted to materials on a hexagonal lattice, and we predict orbital texture switches to be present on square or rectangular lattice materials as well. Through this simple model it is possible to understand the underlying physics of seemingly exotic experimental observations that happen in strong spin orbit coupled materials.

Acknowledgments: This work was funded by NSF DMREF project DMR-1334170 to the University of Colorado and the University of Kentucky. The Advanced Light Source is supported by the Director, Office of Science, Office of Basic Energy Sciences, of the U.S. Department of Energy under Contract No. DE-AC02-05CH11231.  

\bibliography{refs}

\begin{thebibliography}{15}%
\makeatletter
\providecommand \@ifxundefined [1]{%
 \@ifx{#1\undefined}
}%
\providecommand \@ifnum [1]{%
 \ifnum #1\expandafter \@firstoftwo
 \else \expandafter \@secondoftwo
 \fi
}%
\providecommand \@ifx [1]{%
 \ifx #1\expandafter \@firstoftwo
 \else \expandafter \@secondoftwo
 \fi
}%
\providecommand \natexlab [1]{#1}%
\providecommand \enquote  [1]{``#1''}%
\providecommand \bibnamefont  [1]{#1}%
\providecommand \bibfnamefont [1]{#1}%
\providecommand \citenamefont [1]{#1}%
\providecommand \href@noop [0]{\@secondoftwo}%
\providecommand \href [0]{\begingroup \@sanitize@url \@href}%
\providecommand \@href[1]{\@@startlink{#1}\@@href}%
\providecommand \@@href[1]{\endgroup#1\@@endlink}%
\providecommand \@sanitize@url [0]{\catcode `\\12\catcode `\$12\catcode
  `\&12\catcode `\#12\catcode `\^12\catcode `\_12\catcode `\%12\relax}%
\providecommand \@@startlink[1]{}%
\providecommand \@@endlink[0]{}%
\providecommand \url  [0]{\begingroup\@sanitize@url \@url }%
\providecommand \@url [1]{\endgroup\@href {#1}{\urlprefix }}%
\providecommand \urlprefix  [0]{URL }%
\providecommand \Eprint [0]{\href }%
\providecommand \doibase [0]{http://dx.doi.org/}%
\providecommand \selectlanguage [0]{\@gobble}%
\providecommand \bibinfo  [0]{\@secondoftwo}%
\providecommand \bibfield  [0]{\@secondoftwo}%
\providecommand \translation [1]{[#1]}%
\providecommand \BibitemOpen [0]{}%
\providecommand \bibitemStop [0]{}%
\providecommand \bibitemNoStop [0]{.\EOS\space}%
\providecommand \EOS [0]{\spacefactor3000\relax}%
\providecommand \BibitemShut  [1]{\csname bibitem#1\endcsname}%
\let\auto@bib@innerbib\@empty
\bibitem [{\citenamefont {Bawden}\ \emph {et~al.}(2015)\citenamefont {Bawden},
  \citenamefont {Riley}, \citenamefont {Kim}, \citenamefont {Sankar},
  \citenamefont {Monkman}, \citenamefont {Shai}, \citenamefont {Wei},
  \citenamefont {Lochocki}, \citenamefont {Wells}, \citenamefont {Meevasana}
  \emph {et~al.}}]{qt2}%
  \BibitemOpen
  \bibfield  {author} {\bibinfo {author} {\bibfnamefont {L.}~\bibnamefont
  {Bawden}}, \bibinfo {author} {\bibfnamefont {J.~M.}\ \bibnamefont {Riley}},
  \bibinfo {author} {\bibfnamefont {C.~H.}\ \bibnamefont {Kim}}, \bibinfo
  {author} {\bibfnamefont {R.}~\bibnamefont {Sankar}}, \bibinfo {author}
  {\bibfnamefont {E.~J.}\ \bibnamefont {Monkman}}, \bibinfo {author}
  {\bibfnamefont {D.~E.}\ \bibnamefont {Shai}}, \bibinfo {author}
  {\bibfnamefont {H.~I.}\ \bibnamefont {Wei}}, \bibinfo {author} {\bibfnamefont
  {E.~B.}\ \bibnamefont {Lochocki}}, \bibinfo {author} {\bibfnamefont {J.~W.}\
  \bibnamefont {Wells}}, \bibinfo {author} {\bibfnamefont {W.}~\bibnamefont
  {Meevasana}},  \emph {et~al.},\ }\href@noop {} {\bibfield  {journal}
  {\bibinfo  {journal} {Science Advances}\ }\textbf {\bibinfo {volume} {1}},\
  \bibinfo {pages} {e1500495} (\bibinfo {year} {2015})}\BibitemShut {NoStop}%
\bibitem [{\citenamefont {Cao}\ \emph {et~al.}(2013)\citenamefont {Cao},
  \citenamefont {Waugh}, \citenamefont {Zhang}, \citenamefont {Luo},
  \citenamefont {Wang}, \citenamefont {Reber}, \citenamefont {Mo},
  \citenamefont {Xu}, \citenamefont {Yang}, \citenamefont {Schneeloch} \emph
  {et~al.}}]{labios22}%
  \BibitemOpen
  \bibfield  {author} {\bibinfo {author} {\bibfnamefont {Y.}~\bibnamefont
  {Cao}}, \bibinfo {author} {\bibfnamefont {J.}~\bibnamefont {Waugh}}, \bibinfo
  {author} {\bibfnamefont {X.}~\bibnamefont {Zhang}}, \bibinfo {author}
  {\bibfnamefont {J.-W.}\ \bibnamefont {Luo}}, \bibinfo {author} {\bibfnamefont
  {Q.}~\bibnamefont {Wang}}, \bibinfo {author} {\bibfnamefont {T.}~\bibnamefont
  {Reber}}, \bibinfo {author} {\bibfnamefont {S.}~\bibnamefont {Mo}}, \bibinfo
  {author} {\bibfnamefont {Z.}~\bibnamefont {Xu}}, \bibinfo {author}
  {\bibfnamefont {A.}~\bibnamefont {Yang}}, \bibinfo {author} {\bibfnamefont
  {J.}~\bibnamefont {Schneeloch}},  \emph {et~al.},\ }\href@noop {} {\bibfield
  {journal} {\bibinfo  {journal} {Nature Physics}\ }\textbf {\bibinfo {volume}
  {9}},\ \bibinfo {pages} {499} (\bibinfo {year} {2013})}\BibitemShut {NoStop}%
\bibitem [{\citenamefont {Zhang}\ \emph {et~al.}(2013)\citenamefont {Zhang},
  \citenamefont {Liu},\ and\ \citenamefont {Zhang}}]{qt3}%
  \BibitemOpen
  \bibfield  {author} {\bibinfo {author} {\bibfnamefont {H.}~\bibnamefont
  {Zhang}}, \bibinfo {author} {\bibfnamefont {C.-X.}\ \bibnamefont {Liu}}, \
  and\ \bibinfo {author} {\bibfnamefont {S.-C.}\ \bibnamefont {Zhang}},\
  }\href@noop {} {\bibfield  {journal} {\bibinfo  {journal} {Physical review
  letters}\ }\textbf {\bibinfo {volume} {111}},\ \bibinfo {pages} {066801}
  (\bibinfo {year} {2013})}\BibitemShut {NoStop}%
\bibitem [{\citenamefont {Qi}\ and\ \citenamefont
  {Zhang}(2011)}]{qi2011topological}%
  \BibitemOpen
  \bibfield  {author} {\bibinfo {author} {\bibfnamefont {X.-L.}\ \bibnamefont
  {Qi}}\ and\ \bibinfo {author} {\bibfnamefont {S.-C.}\ \bibnamefont {Zhang}},\
  }\href@noop {} {\bibfield  {journal} {\bibinfo  {journal} {Reviews of Modern
  Physics}\ }\textbf {\bibinfo {volume} {83}},\ \bibinfo {pages} {1057}
  (\bibinfo {year} {2011})}\BibitemShut {NoStop}%
\bibitem [{\citenamefont {Zhu}\ \emph {et~al.}(2013)\citenamefont {Zhu},
  \citenamefont {Veenstra}, \citenamefont {Levy}, \citenamefont {Ubaldini},
  \citenamefont {Syers}, \citenamefont {Butch}, \citenamefont {Paglione},
  \citenamefont {Haverkort}, \citenamefont {Elfimov},\ and\ \citenamefont
  {Damascelli}}]{qt7}%
  \BibitemOpen
  \bibfield  {author} {\bibinfo {author} {\bibfnamefont {Z.-H.}\ \bibnamefont
  {Zhu}}, \bibinfo {author} {\bibfnamefont {C.}~\bibnamefont {Veenstra}},
  \bibinfo {author} {\bibfnamefont {G.}~\bibnamefont {Levy}}, \bibinfo {author}
  {\bibfnamefont {A.}~\bibnamefont {Ubaldini}}, \bibinfo {author}
  {\bibfnamefont {P.}~\bibnamefont {Syers}}, \bibinfo {author} {\bibfnamefont
  {N.}~\bibnamefont {Butch}}, \bibinfo {author} {\bibfnamefont
  {J.}~\bibnamefont {Paglione}}, \bibinfo {author} {\bibfnamefont
  {M.}~\bibnamefont {Haverkort}}, \bibinfo {author} {\bibfnamefont
  {I.}~\bibnamefont {Elfimov}}, \ and\ \bibinfo {author} {\bibfnamefont
  {A.}~\bibnamefont {Damascelli}},\ }\href@noop {} {\bibfield  {journal}
  {\bibinfo  {journal} {Physical review letters}\ }\textbf {\bibinfo {volume}
  {110}},\ \bibinfo {pages} {216401} (\bibinfo {year} {2013})}\BibitemShut
  {NoStop}%
\bibitem [{\citenamefont {Cao}\ \emph {et~al.}(2012)\citenamefont {Cao},
  \citenamefont {Waugh}, \citenamefont {Plumb}, \citenamefont {Reber},
  \citenamefont {Parham}, \citenamefont {Landolt}, \citenamefont {Xu},
  \citenamefont {Yang}, \citenamefont {Schneeloch}, \citenamefont {Gu} \emph
  {et~al.}}]{qt8}%
  \BibitemOpen
  \bibfield  {author} {\bibinfo {author} {\bibfnamefont {Y.}~\bibnamefont
  {Cao}}, \bibinfo {author} {\bibfnamefont {J.}~\bibnamefont {Waugh}}, \bibinfo
  {author} {\bibfnamefont {N.}~\bibnamefont {Plumb}}, \bibinfo {author}
  {\bibfnamefont {T.}~\bibnamefont {Reber}}, \bibinfo {author} {\bibfnamefont
  {S.}~\bibnamefont {Parham}}, \bibinfo {author} {\bibfnamefont
  {G.}~\bibnamefont {Landolt}}, \bibinfo {author} {\bibfnamefont
  {Z.}~\bibnamefont {Xu}}, \bibinfo {author} {\bibfnamefont {A.}~\bibnamefont
  {Yang}}, \bibinfo {author} {\bibfnamefont {J.}~\bibnamefont {Schneeloch}},
  \bibinfo {author} {\bibfnamefont {G.}~\bibnamefont {Gu}},  \emph {et~al.},\
  }\href@noop {} {\bibfield  {journal} {\bibinfo  {journal} {arXiv preprint
  arXiv:1211.5998}\ } (\bibinfo {year} {2012})}\BibitemShut {NoStop}%
\bibitem [{\citenamefont {Zhu}\ \emph {et~al.}(2014)\citenamefont {Zhu},
  \citenamefont {Veenstra}, \citenamefont {Zhdanovich}, \citenamefont
  {Schneider}, \citenamefont {Okuda}, \citenamefont {Miyamoto}, \citenamefont
  {Zhu}, \citenamefont {Namatame}, \citenamefont {Taniguchi}, \citenamefont
  {Haverkort} \emph {et~al.}}]{qt9}%
  \BibitemOpen
  \bibfield  {author} {\bibinfo {author} {\bibfnamefont {Z.-H.}\ \bibnamefont
  {Zhu}}, \bibinfo {author} {\bibfnamefont {C.}~\bibnamefont {Veenstra}},
  \bibinfo {author} {\bibfnamefont {S.}~\bibnamefont {Zhdanovich}}, \bibinfo
  {author} {\bibfnamefont {M.}~\bibnamefont {Schneider}}, \bibinfo {author}
  {\bibfnamefont {T.}~\bibnamefont {Okuda}}, \bibinfo {author} {\bibfnamefont
  {K.}~\bibnamefont {Miyamoto}}, \bibinfo {author} {\bibfnamefont {S.-Y.}\
  \bibnamefont {Zhu}}, \bibinfo {author} {\bibfnamefont {H.}~\bibnamefont
  {Namatame}}, \bibinfo {author} {\bibfnamefont {M.}~\bibnamefont {Taniguchi}},
  \bibinfo {author} {\bibfnamefont {M.}~\bibnamefont {Haverkort}},  \emph
  {et~al.},\ }\href@noop {} {\bibfield  {journal} {\bibinfo  {journal}
  {Physical review letters}\ }\textbf {\bibinfo {volume} {112}},\ \bibinfo
  {pages} {076802} (\bibinfo {year} {2014})}\BibitemShut {NoStop}%
\bibitem [{\citenamefont {Rashba}(1960)}]{labios3}%
  \BibitemOpen
  \bibfield  {author} {\bibinfo {author} {\bibfnamefont {E.}~\bibnamefont
  {Rashba}},\ }\href@noop {} {\bibfield  {journal} {\bibinfo  {journal} {Soviet
  Physics-Solid State}\ }\textbf {\bibinfo {volume} {2}},\ \bibinfo {pages}
  {1109} (\bibinfo {year} {1960})}\BibitemShut {NoStop}%
\bibitem [{\citenamefont {Petersen}\ and\ \citenamefont
  {Hedeg{\aa}rd}(2000)}]{qt6}%
  \BibitemOpen
  \bibfield  {author} {\bibinfo {author} {\bibfnamefont {L.}~\bibnamefont
  {Petersen}}\ and\ \bibinfo {author} {\bibfnamefont {P.}~\bibnamefont
  {Hedeg{\aa}rd}},\ }\href@noop {} {\bibfield  {journal} {\bibinfo  {journal}
  {Surface science}\ }\textbf {\bibinfo {volume} {459}},\ \bibinfo {pages} {49}
  (\bibinfo {year} {2000})}\BibitemShut {NoStop}%
\bibitem [{\citenamefont {Park}\ \emph {et~al.}(2011)\citenamefont {Park},
  \citenamefont {Kim}, \citenamefont {Yu}, \citenamefont {Han},\ and\
  \citenamefont {Kim}}]{qt12}%
  \BibitemOpen
  \bibfield  {author} {\bibinfo {author} {\bibfnamefont {S.~R.}\ \bibnamefont
  {Park}}, \bibinfo {author} {\bibfnamefont {C.~H.}\ \bibnamefont {Kim}},
  \bibinfo {author} {\bibfnamefont {J.}~\bibnamefont {Yu}}, \bibinfo {author}
  {\bibfnamefont {J.~H.}\ \bibnamefont {Han}}, \ and\ \bibinfo {author}
  {\bibfnamefont {C.}~\bibnamefont {Kim}},\ }\href@noop {} {\bibfield
  {journal} {\bibinfo  {journal} {Physical review letters}\ }\textbf {\bibinfo
  {volume} {107}},\ \bibinfo {pages} {156803} (\bibinfo {year}
  {2011})}\BibitemShut {NoStop}%
\bibitem [{\citenamefont {Liu}\ \emph {et~al.}(2016)\citenamefont {Liu},
  \citenamefont {Waugh}, \citenamefont {Dessau},\ and\ \citenamefont
  {Zunger}}]{liu2016orbital}%
  \BibitemOpen
  \bibfield  {author} {\bibinfo {author} {\bibfnamefont {Q.}~\bibnamefont
  {Liu}}, \bibinfo {author} {\bibfnamefont {X.~Z.~J.}\ \bibnamefont {Waugh}},
  \bibinfo {author} {\bibfnamefont {D.}~\bibnamefont {Dessau}}, \ and\ \bibinfo
  {author} {\bibfnamefont {A.}~\bibnamefont {Zunger}},\ }\href@noop {}
  {\bibfield  {journal} {\bibinfo  {journal} {arXiv preprint arXiv:1607.01692}\
  } (\bibinfo {year} {2016})}\BibitemShut {NoStop}%
\bibitem [{\citenamefont {Dresselhaus}(1955)}]{labios4}%
  \BibitemOpen
  \bibfield  {author} {\bibinfo {author} {\bibfnamefont {G.}~\bibnamefont
  {Dresselhaus}},\ }\href {\doibase 10.1103/PhysRev.100.580} {\bibfield
  {journal} {\bibinfo  {journal} {Phys. Rev.}\ }\textbf {\bibinfo {volume}
  {100}},\ \bibinfo {pages} {580} (\bibinfo {year} {1955})}\BibitemShut
  {NoStop}%
\bibitem [{\citenamefont {Riley}\ \emph {et~al.}(2014)\citenamefont {Riley},
  \citenamefont {Mazzola}, \citenamefont {Dendzik}, \citenamefont {Michiardi},
  \citenamefont {Takayama}, \citenamefont {Bawden}, \citenamefont
  {Graner{\o}d}, \citenamefont {Leandersson}, \citenamefont {Balasubramanian},
  \citenamefont {Hoesch} \emph {et~al.}}]{qt10}%
  \BibitemOpen
  \bibfield  {author} {\bibinfo {author} {\bibfnamefont {J.}~\bibnamefont
  {Riley}}, \bibinfo {author} {\bibfnamefont {F.}~\bibnamefont {Mazzola}},
  \bibinfo {author} {\bibfnamefont {M.}~\bibnamefont {Dendzik}}, \bibinfo
  {author} {\bibfnamefont {M.}~\bibnamefont {Michiardi}}, \bibinfo {author}
  {\bibfnamefont {T.}~\bibnamefont {Takayama}}, \bibinfo {author}
  {\bibfnamefont {L.}~\bibnamefont {Bawden}}, \bibinfo {author} {\bibfnamefont
  {C.}~\bibnamefont {Graner{\o}d}}, \bibinfo {author} {\bibfnamefont
  {M.}~\bibnamefont {Leandersson}}, \bibinfo {author} {\bibfnamefont
  {T.}~\bibnamefont {Balasubramanian}}, \bibinfo {author} {\bibfnamefont
  {M.}~\bibnamefont {Hoesch}},  \emph {et~al.},\ }\href@noop {} {\bibfield
  {journal} {\bibinfo  {journal} {Nature Physics}\ }\textbf {\bibinfo {volume}
  {10}},\ \bibinfo {pages} {835} (\bibinfo {year} {2014})}\BibitemShut
  {NoStop}%
\bibitem [{\citenamefont {Zhang}\ \emph {et~al.}(2014)\citenamefont {Zhang},
  \citenamefont {Liu}, \citenamefont {Luo}, \citenamefont {Freeman},\ and\
  \citenamefont {Zunger}}]{labios2}%
  \BibitemOpen
  \bibfield  {author} {\bibinfo {author} {\bibfnamefont {X.}~\bibnamefont
  {Zhang}}, \bibinfo {author} {\bibfnamefont {Q.}~\bibnamefont {Liu}}, \bibinfo
  {author} {\bibfnamefont {J.-W.}\ \bibnamefont {Luo}}, \bibinfo {author}
  {\bibfnamefont {A.~J.}\ \bibnamefont {Freeman}}, \ and\ \bibinfo {author}
  {\bibfnamefont {A.}~\bibnamefont {Zunger}},\ }\href@noop {} {\bibfield
  {journal} {\bibinfo  {journal} {Nature Physics}\ }\textbf {\bibinfo {volume}
  {10}},\ \bibinfo {pages} {387} (\bibinfo {year} {2014})}\BibitemShut
  {NoStop}%
\bibitem [{\citenamefont {Mizuguchi}\ \emph {et~al.}(2012)\citenamefont
  {Mizuguchi}, \citenamefont {Demura}, \citenamefont {Deguchi}, \citenamefont
  {Takano}, \citenamefont {Fujihisa}, \citenamefont {Gotoh}, \citenamefont
  {Izawa},\ and\ \citenamefont {Miura}}]{labios6}%
  \BibitemOpen
  \bibfield  {author} {\bibinfo {author} {\bibfnamefont {Y.}~\bibnamefont
  {Mizuguchi}}, \bibinfo {author} {\bibfnamefont {S.}~\bibnamefont {Demura}},
  \bibinfo {author} {\bibfnamefont {K.}~\bibnamefont {Deguchi}}, \bibinfo
  {author} {\bibfnamefont {Y.}~\bibnamefont {Takano}}, \bibinfo {author}
  {\bibfnamefont {H.}~\bibnamefont {Fujihisa}}, \bibinfo {author}
  {\bibfnamefont {Y.}~\bibnamefont {Gotoh}}, \bibinfo {author} {\bibfnamefont
  {H.}~\bibnamefont {Izawa}}, \ and\ \bibinfo {author} {\bibfnamefont
  {O.}~\bibnamefont {Miura}},\ }\href@noop {} {\bibfield  {journal} {\bibinfo
  {journal} {Journal of the Physical Society of Japan}\ }\textbf {\bibinfo
  {volume} {81}},\ \bibinfo {pages} {114725} (\bibinfo {year}
  {2012})}\BibitemShut {NoStop}%
\end{thebibliography}%


\end{document}